\documentclass[english,aps,manuscript]{revtex4}
\usepackage[T1]{fontenc}
\usepackage[latin9]{inputenc}
\usepackage{float}
\usepackage{textcomp}
\usepackage{amsmath}
\usepackage{amssymb}
\usepackage{graphicx}
\usepackage{esint}

\makeatletter

\newcommand{\lyxmathsym}[1]{\ifmmode\begingroup\def\b@ld{bold}
  \text{\ifx\math@version\b@ld\bfseries\fi#1}\endgroup\else#1\fi}

\@ifundefined{textcolor}{}
{%
 \definecolor{BLACK}{gray}{0}
 \definecolor{WHITE}{gray}{1}
 \definecolor{RED}{rgb}{1,0,0}
 \definecolor{GREEN}{rgb}{0,1,0}
 \definecolor{BLUE}{rgb}{0,0,1}
 \definecolor{CYAN}{cmyk}{1,0,0,0}
 \definecolor{MAGENTA}{cmyk}{0,1,0,0}
 \definecolor{YELLOW}{cmyk}{0,0,1,0}
 }

\@ifundefined{showcaptionsetup}{}{%
 \PassOptionsToPackage{caption=false}{subfig}}
\usepackage{subfig}
\makeatother

\usepackage{babel}
\begin{document}

\title{Describing Spin-Selective Reactions of Radical Pairs Using Quantum
Jump Approaches}

\author{Alpha Lee}

\date{10 Oct 2011}

\altaffiliation{Department of Chemistry, University of Oxford, Physical and Theoretical Chemistry Laboratory, South Parks Road, Oxford OX1 3QZ}
\altaffiliation{Department of Chemistry, Faculty of Natural Sciences, Imperial College London, London SW7 2AZ}

\email{alpha.lee09@imperial.ac.uk}

\begin{abstract}
Recently, spin-selective radical pair reactions have been studied
using concepts from quantum measurement theory. In this Article, we
show that the approach taken by Kominis (Physical Review E, 83, 2011,
056118) leads to erroneous results due to a problematic treatment
of quantum jumps. Correct consideration of quantum jumps leads to
either the traditional master equation or the Jones-Hore master equation. 
\end{abstract}
\maketitle

\section{Introduction}

Spin-selective radical\textendash{}ion pair reactions are at the core
of spin chemistry. The phenomenological master equation \cite{HABERKORN:1976ss}
(1) is often used to model spin-selective reactions. 
\begin{equation}
\frac{d\rho}{dt}=-i[H,\rho]-\frac{k_{S}}{2}\{Q_{S},\rho\}-\frac{k_{T}}{2}\{Q_{T},\rho\}
\end{equation}

where $\rho$ is the density matrix of the reactants, $\left\{ \right\} $
denotes the anti-commutator, $H$ is the Hamiltonian describing unitary
evolution of the radical pair, $k_{S}$ and $k_{T}$ are the rates
of reaction through singlet and triplet channels respectively, and
$Q_{S}$ and $Q_{T}$ are singlet and triplet projection operators
respectively. Recently, based on quantum measurement theory, Kominis
\cite{Kominis:2009ek,Kominis:2010gf} derived equation (2) and quantum
jump equations (3) and (4) 
\begin{equation}
\frac{d\rho_{nr}}{dt}=-i[H,\rho]-\frac{k_{S}+k_{T}}{2}(\rho Q_{S}+Q_{S}\rho-2Q_{S}\rho Q_{S})
\end{equation}
 $\rho_{nr}$ describes the state of the radical pairs before recombination.
Equation (2) is derived by making the analogy between spin-selective
chemical reaction and electron tunneling in a quantum dot. The quantum
non-demolition measurement of the quantum dot causes the wavefunction
to collapse periodically, thus pure decoherence without energy dissipation
is observed. Kominis then introduced quantum jumps (3) and (4), where
$p_{S}$ is the probability of reacting down the singlet channel and
$p_{T}$ is the probability of reacting down the triplet channel,
to represent chemical reaction. 

\begin{equation}
p_{S}=k_{S}\left\langle Q_{S}\right\rangle dt
\end{equation}
\begin{equation}
p_{T}=k_{T}\left\langle Q_{T}\right\rangle dt
\end{equation}

This an erroneous application of the Bohr-Einstein quantum jump \cite{Carmichael:1997uk}
and leads to misinterpretation of the {}``no jump'' event \cite{MOLMER:1993rr}
which will be analysed below. Subsequently, Kominis \cite{Kominis:2011sf}
introduced a revised master equation based on a phenomenological interpolation
between the {}``maximal coherence case'' and the {}``minimal coherence
case'' without correcting the underlying physics in (3) and (4).
Rephrasing the argument presented, singlet-triplet coherence is measured
by the parameter $\rho_{coh}$.

\begin{equation}
\rho_{coh}=\frac{Tr\{\rho_{ST}\rho_{TS}\}}{Tr\{\rho_{SS}\}Tr\{\rho_{TT}\}}
\end{equation}
where $\rho_{ST}=Q_{S}\rho Q_{T}$, $\rho_{TS}=Q_{T}\rho Q_{S}$,
$\rho_{SS}=Q_{S}\rho Q_{S}$ and $\rho_{TT}=Q_{T}\rho Q_{T}$ . The
evolution of the density matrix is given by 

\begin{equation}
\frac{d\rho}{dt}=-i[H,\rho]-\frac{k_{S}+k_{T}}{2}(\rho Q_{S}+Q_{S}\rho-2Q_{S}\rho Q_{S})-(1-\rho_{coh})\frac{d\rho_{incoh}}{dt}-\rho_{coh}\frac{d\rho_{coh}}{dt}
\end{equation}

\begin{equation}
d\rho_{incoh}=k_{S}dtQ_{S}\rho Q_{S}+k_{T}dtQ_{T}\rho Q_{T}
\end{equation}
\begin{equation}
d\rho_{coh}=(k_{S}Tr\{Q_{S}\rho\}+k_{T}Tr\{Q_{T}\rho\})dt\frac{\rho}{Tr\{\rho\}}
\end{equation}

The interpolation parameter $\rho_{coh}$ is introduced {}``by hand''
and the limiting cases of the resulting equation will be commented
on below. This Article will also show how consistent derivation of
the master equation points towards (9), the Jones-Hore master equation
\cite{Jones:2010lk} or the phenomenological master equation. 

\begin{equation}
\frac{d\rho}{dt}=-i[H,\rho]-\frac{k_{S}}{2}\{Q_{S},\rho\}-\frac{k_{T}}{2}\{Q_{T},\rho\}-\frac{k_{S}+k_{T}}{2}(\rho Q_{S}+Q_{S}\rho-2Q_{S}\rho Q_{S})
\end{equation}

A microscopic derivation of (1) has been reported by Ivanov et al.
\cite{Ivanov:2010ul}. Appendix A contains the full microscopic derivation
of (9).

\section{Comment on the Kominis master equation}

\subsection{Problem with {}``no jump'' events }

The quantum jump equations (3) and (4) fail to capture the physical
significance of a no-jump event. In a qualitative sense, a no-jump
event represents either the wavefunction is still in $\{S,T\}$ subspace
or, importantly, the fact that the radical pair has already reacted
\cite{Wiseman:1996hw,Plenio:1998wn}. If one starts observing the
system at time $t$ and does not see the system executing a quantum
jump after a very long time, one should conclude that the system has
already jumped at some prior time before the observation rather then
believing that the system is yet to jump. To capture this, a non-Hermitian
term must be added to the Hamiltonian 
\begin{equation}
H_{eff}=H-i\frac{k_{S}}{2}\left|S\left\rangle \right\langle S\right|-i\frac{k_{T}}{2}\left|T\left\rangle \right\langle T\right|
\end{equation}

Putting into (2) the non-Hermitian Hamiltonian (10), which corresponds
to quantum jump equations (3) and (4), and only considering the S-T
subspace

\begin{eqnarray*}
\frac{d\rho}{dt} & = & \frac{d\rho_{reaction}}{dt}+\frac{d\rho_{nr}}{dt}\\
 & = & -i[H-i\frac{k_{S}}{2}\left|S\left\rangle \right\langle S\right|-i\frac{k_{T}}{2}\left|T\left\rangle \right\langle T\right|,\rho]-\frac{k_{S}+k_{T}}{2}(\rho Q_{S}+Q_{S}\rho-2Q_{S}\rho Q_{S})\\
 & = & -i[H,\rho]-\frac{k_{S}}{2}\{Q_{S},\rho\}-\frac{k_{T}}{2}\{Q_{T},\rho\}-\frac{k_{S}+k_{T}}{2}(\rho Q_{S}+Q_{S}\rho-2Q_{S}\rho Q_{S})
\end{eqnarray*}
and the Jones-Hore master equation \cite{Jones:2010lk} is recovered.

The role of the non-Hermitian Hamiltonian can be seen more clearly
when one writes the jump-free evolution of the system explicitly.
When $\left\Vert H_{eff}\right\Vert dt\ll1$ 
\begin{eqnarray}
\left|\psi(t+dt)\right\rangle  & = & e^{-iH_{eff}dt}\left|\psi(t)\right\rangle \nonumber \\
 & \approx & (1-iH_{eff}dt)\left|\psi(t)\right\rangle 
\end{eqnarray}

ignoring second order terms in $dt$

\begin{eqnarray}
\left|\psi(t+dt)\right|^{2} & = & \left\langle \psi(t)\right|(1+iH_{eff}^{+}dt)(1-iH_{eff}dt)\left|\psi(t)\right\rangle \nonumber \\
 & = & 1-\delta p
\end{eqnarray}
where $\delta p$ is the decrease in the norm of the wavefunction
which is compensated by the jump to products. Identifying $\delta p$
as the probability of a quantum jump
\begin{eqnarray}
\delta p & = & idt\left\langle \psi(t)\right|H_{eff}-H_{eff}^{+}\left|\psi(t)\right\rangle \nonumber \\
 & = & dt(k_{S}\left\langle \psi(t)\right|Q_{S}\left|\psi(t)\right\rangle +k_{T}\left\langle \psi(t)\right|Q_{T}\left|\psi(t)\right\rangle )\nonumber \\
 & = & k_{S}dt\left\langle Q_{S}\right\rangle +k_{T}dt\left\langle Q_{T}\right\rangle 
\end{eqnarray}

This shows qualitatively that although the jump operators {}``fill
up'' the states representing the reaction products ($S_{0}$ and
$T_{0}$), the populations of the radical pair $S$ and $T$ states
are only {}``removed'' by the non-Hermitian Hamiltonian. 

From an algebraic perspective, any completely positive Markovian evolution
can be written in the Lindblad form
\begin{eqnarray}
\frac{d\rho}{dt} & = & -i[H,\rho]+D(\rho)\nonumber \\
 & = & -i[H,\rho]+\sum_{i}L_{i}\rho L_{i}^{+}-\frac{1}{2}\{L_{i}^{+}L_{i},\rho\}
\end{eqnarray}
where $L_{i}$ are Lindblad operators. The formal solution to (14)
can be written as 
\begin{equation}
\rho(t)=e^{\hat{\hat{L}}t}\rho(0)
\end{equation}
$\hat{\hat{L}}$ is a superoperator that can be written trivially
in terms of the jump superoperator $\hat{\hat{J}}$ 
\begin{equation}
\hat{\hat{L}}=\hat{\hat{J}}+\hat{\hat{L}}-\hat{\hat{J}}
\end{equation}

using an identity for superoperators \cite{carmichael} 
\begin{equation}
e^{(\hat{\hat{a}}+\hat{\hat{b}})x}=\sum_{k=0}^{\infty}\int_{0}^{x}dx_{k}\int_{0}^{x_{k}}dx_{k-1}...\int_{0}^{x_{2}}dx_{1}e^{\hat{\hat{a}}(x-x_{k})}\hat{\hat{b}}e^{\hat{\hat{a}}(x_{k}-x_{k-1})}\hat{\hat{b}}...\hat{\hat{b}}e^{\hat{\hat{a}}x_{1}}
\end{equation}
\begin{equation}
e^{\hat{(\hat{J}}+\hat{\hat{L}}-\hat{\hat{J}})t}\rho(0)=\sum_{k=0}^{\infty}\int_{0}^{t}dt_{k}\int_{0}^{t_{k}}dt_{k-1}...\int_{0}^{t_{2}}dt_{1}e^{\hat{(\hat{L}}-\hat{\hat{J}})(t-t_{k})}\hat{\hat{J}}e^{(\hat{\hat{L}}-\hat{\hat{J}})(t_{k}-t_{k-1})}\hat{\hat{J}}...\hat{\hat{J}}e^{(\hat{\hat{L}}-\hat{\hat{J}})t_{1}}\rho(0)
\end{equation}

Reading (18) from right to left, $\hat{\hat{L}}-\hat{\hat{J}}$ can
be interpreted as the {}``between jump'' evolution of the system
and at $t_{1}$ the system experiences a first jump, followed by a
period of between jump evolution and so on. $\hat{\hat{J}}$, the
jump operator, can be identified with terms bilinear in $\rho$. 
\begin{equation}
\hat{\hat{J}}=\sum_{i}L_{i}\bullet L_{i}^{+}
\end{equation}
\begin{equation}
\hat{\hat{L}}=-iH\bullet+\bullet iH+\sum_{i}\left(L_{i}\bullet L_{i}^{+}-\frac{1}{2}L_{i}^{+}L_{i}\bullet-\frac{1}{2}\bullet L_{i}^{+}L_{i}\right)
\end{equation}

\begin{eqnarray}
\hat{\hat{L}}-\hat{\hat{J}} & = & -iH\bullet+\bullet iH-\sum_{i}\left(\frac{1}{2}L_{i}^{+}L_{i}\bullet+\frac{1}{2}\bullet L_{i}^{+}L_{i}\right)\nonumber \\
 & = & -i(H\bullet-\bullet H-\sum_{i}\left(\frac{i}{2}L_{i}^{+}L_{i}\bullet+\frac{i}{2}\bullet L_{i}^{+}L_{i}\right))\nonumber \\
 & = & -i[H-\frac{i}{2}\sum_{i}L_{i}^{+}L_{i},\bullet]
\end{eqnarray}

Identifying the fact that if a quantum jump is not executed between
time interval $t$ and $t+dt$, $e^{(\hat{\hat{L}}-\hat{\hat{J}})dt}$
propagates the state $\rho(t)\rightarrow\rho(t+dt)$. 

\begin{equation}
\left|\psi(t+dt)\right\rangle =e^{-iH_{eff}dt}\left|\psi(t)\right\rangle 
\end{equation}

where 
\begin{equation}
H_{eff}=H-\frac{i}{2}\sum_{i}L_{i}^{+}L_{i}
\end{equation}

Substituting the jump operators corresponding to (3) and (4), which
will be discussed below, into (23) we obtained (10). This shows clearly
how the non-Hermitian Hamiltonian is central to the quantum jump formalism.
It is Kominis's neglect of this quantity that led to erroneous results.

\subsection{Expression for $d\rho_{incoh}$}

If there are no coherences, $\rho=Q_{S}\rho Q_{S}+Q_{T}\rho Q_{T}$
and equation (7) follows trivially. Writing (6) in the limit $\rho_{coh}=0$

\begin{eqnarray*}
\frac{d\rho}{dt} & = & -i[H,\rho]-\frac{k_{S}+k_{T}}{2}(\rho Q_{S}+Q_{S}\rho-2Q_{S}\rho Q_{S})-k_{S}Q_{S}\rho Q_{S}-k_{T}Q_{T}\rho Q_{T}\\
 & = & -i[H,\rho]-\frac{k_{S}}{2}\{Q_{S},\rho\}-\frac{k_{T}}{2}\{Q_{T},\rho\}
\end{eqnarray*}
The Jones-Hore equation (9) also reduces to same equation.

\subsection{Expression for\textmd{ $d\rho_{coh}$}}

The expression for $d\rho_{coh}$ suggests that the total density
matrix is removed at the combined rate at which the reactants are
transformed to product. This approach is no longer state selective
as we know that $\left|S\right\rangle \rightarrow\left|T_{0}\right\rangle $
and $\left|T\right\rangle \rightarrow\left|S_{0}\right\rangle $ transitions
are forbidden. Hence a differential rate between singlet recombination
and triplet recombination should manifest itself in different rates
at which the singlet and triplet states are depopulated. This suggests
that the expression for $d\rho_{coh}$ is correct only if $k_{S}=k_{T}$.
In the limit $ $$\rho_{coh}=1$ and $k_{S}=k_{T}=k$, (6) reads
\begin{equation}
\frac{d\rho}{dt}=-i[H,\rho]-k(\rho Q_{S}+Q_{S}\rho-2Q_{S}\rho Q_{S})-(kTr\{Q_{S}\rho\}+kTr\{Q_{T}\rho\})\frac{\rho}{Tr\{\rho\}}
\end{equation}
Knowing that 
\begin{equation}
Tr\{\rho\}=Tr\{Q_{S}\rho\}+Tr\{Q_{T}\rho\}
\end{equation}

we find

\begin{equation}
\frac{d\rho}{dt}=-i[H,\rho]-k(\rho Q_{S}+Q_{S}\rho-2Q_{S}\rho Q_{S})-k\rho
\end{equation}
(26) is exactly equal to the Jones-Hore equation (9) in the same limit.

\subsection{Unphysical Prediction of the Kominis Master Equation }

The Kominis master equation predicts that in the absence of singlet-triplet
interconversion and $k_{T}=0$, starting from a totally coherent mixture
of $S$ and $T$ one will have triplet population equal to 0.25 after
reaction \cite{Kominis:2011sf}. 

Kominis attempted to defend the unexpected fall in $\left\langle Q_{T}\right\rangle $
by analysing a single molecule trajectory. Paraphrasing his argument,
at $t=0$, the radical pair can either react with probability $p_{r}=k_{S}dt\left\langle Q_{S}\right\rangle =\frac{k_{S}dt}{2}$
or not react with probability $p_{nr}=1-\frac{k_{S}dt}{2}$. Conditioned
on the fact that the radical pairs do not react, measurement at rate
$\frac{k_{S}}{2}$ causes the singlet projection to occur with probability
$q_{S}=\frac{k_{S}\left\langle Q_{S}\right\rangle dt}{2}=\frac{k_{S}dt}{4}$
, triplet projection to occur with with probability $q_{T}=\frac{k_{S}\left\langle Q_{T}\right\rangle dt}{2}=\frac{k_{S}dt}{4}$
and the probability of no projection $q_{0}=1-\frac{k_{S}dt}{2}$.
As pure singlet will react eventually and pure triplet will never
react, summing up pure singlet produced by the measurement and singlet
product will give the total singlet yield. 
\begin{equation}
Y_{S}=(p_{r}+p_{nr}q_{S})+p_{nr}q_{0}(p_{r}+p_{nr}q_{S})+(p_{nr}q_{0})^{2}(p_{r}+p_{nr}q_{S})\ldots
\end{equation}
\begin{equation}
p_{nr}q_{0}=\left(1-\frac{k_{S}dt}{2}\right)^{2}\approx1-k_{S}dt
\end{equation}
\begin{equation}
p_{r}+p_{nr}q_{S}=\frac{k_{S}dt}{2}+\left(1-\frac{k_{S}dt}{2}\right)\frac{k_{S}}{4}dt\thickapprox\frac{3k_{S}dt}{4}
\end{equation}

By summing $Y_{S}$, Kominis obtains 
\begin{equation}
Y_{S}=\frac{3k_{S}dt}{4}\sum_{n=0}^{\infty}(1-k_{S}dt)^{n}=\frac{3}{4}
\end{equation}

The error in this summing procedure is the fact that the whole density
matrix is removed with probability $p_{r}=k_{S}dt\left\langle Q_{S}\right\rangle $
to form the singlet product in the first step. This leads to non-conservation
of spin angular momentum. 

Kominis attempted to circumvent this lack of conservation of spin
angular momentum by invoking a time averaged $\rho_{coh}$. He argues
that 
\begin{equation}
\rho_{coh}(t)=\frac{\text{\textlangle\textlangle}Tr\{\rho_{ST}(t)\rho_{TS}(t+\text{\ensuremath{\tau}})\}\lyxmathsym{\textrangle\textrangle}}{Tr\{\rho_{SS}\}Tr\{\rho_{TT}\}}
\end{equation}

\begin{equation}
\rho_{TS}(t+\tau)=e^{-iH\tau}\rho_{TS}(t)e^{iH\tau}
\end{equation}

where $\left\langle \left\langle \ldots\right\rangle \right\rangle $
indicates time average over $\tau$, with $\tau$ being larger than
the inverse S-T energy difference and smaller than the characteristic
time-scale of the reaction. Assuming the S-T energy separation is
$J$, as $e^{-iJt}$ rotates rapidly around the complex plane

\begin{equation}
\rho_{coh}(t)=\frac{\text{\textlangle\textlangle}Tr\{\rho_{ST}(t)e^{-iJ\tau}\rho_{TS}(t)e^{iJ\tau}\}\text{\textrangle\textrangle}}{Tr\{\rho_{SS}\}Tr\{\rho_{TT}\}}\approx0
\end{equation}

However, (31) is physically questionable. There is no necessary relationship
between the S-T energy difference and the characteristic time-scale
of the reaction, both are system-dependent parameters. 

Furthermore, the validity of (32) is dubious. By integrating the Liouville-von
Neuman equation 
\begin{equation}
\rho(t+\tau)=e^{-iH\tau}\rho(t)e^{iH\tau}
\end{equation}

\begin{equation}
\rho_{TS}(t+\tau)=Q_{T}e^{-iH\tau}\rho(t)e^{iH\tau}Q_{S}
\end{equation}
only when $[H,Q_{S}]=0$ and $[H,Q_{T}]=0$, i.e. in the absence of
singlet-triplet interconversion, can the order of the exponentials
be reversed, so that (32) is recovered.

\subsection{Radical-ion-pair reactions and the optical double slit}

The introduction of $\rho_{coh}$ and the analogy with the optical
double slit experiment suggests that there is singlet interference
to triplet product and triplet interference to singlet product. In
an optical double slit, the field incident on the screen at position
$\overrightarrow{r}$ at time $t$ is the superposition of the fields
from the two slits
\begin{equation}
E^{(+)}(\overrightarrow{r},t)=E_{1}^{(+)}(\overrightarrow{r},t)+E_{2}^{(+)}(\overrightarrow{r},t)
\end{equation}

This is because the two slits are physically identical in an optical
double slit experiment and a single photon can be diffracted by both
slits. In a chemical reaction, spin angular momentum conservation
demands that only singlet reactant can enter the {}``singlet reaction
slit'' and only triplet reactant can enter the {}``triplet reaction
slit''. A superposition state cannot react, the coherence is destroyed
as the starting state for the jump is selected from among the stationary
states represented in the superposition.

\section{Consistent Derivation of Master Equation Using Quantum Jump Approaches}

\subsection{Consistent Derivation of Jones-Hore Master Equation }

Treating quantum measurement and chemical reaction as quantum jump
processes gives a physical interpretation of the Jones-Hore master
equation. Using jump operators (37) - (40) , the master equation (9)
can be recovered. 
\begin{equation}
J_{1}=\left|S_{0}\left\rangle \right\langle S\right|
\end{equation}
\begin{equation}
J_{2}=\left|T_{0}\left\rangle \right\langle T\right|
\end{equation}
\begin{equation}
J_{3}=\left|T\left\rangle \right\langle T\right|
\end{equation}
\begin{equation}
J_{4}=\left|S\left\rangle \right\langle S\right|
\end{equation}
(37) and (38) are the same as (3) and (4) and correspond to quantum
jumps to singlet and triplet product respectively. (39) and (40) correspond
to a strong measurement of the system, the same intuition that lead
to equation (2). The evolution of the wavefunction, $\left|\psi(t)\right\rangle \rightarrow\left|\psi(t+dt)\right\rangle $
is described by
\begin{eqnarray}
\left|\psi(t+dt)\right\rangle =\frac{e^{-iH_{eff}dt}\left|\psi(t)\right\rangle }{\sqrt{\left\langle \Psi\right|e^{iH_{eff}^{+}dt}e^{-iH_{eff}dt}\left|\Psi\right\rangle }} &  & p=1-k_{S}dt-k_{T}dt\nonumber \\
\left|\psi(t+dt)\right\rangle =\left|S_{0}\right\rangle =\frac{J_{1}\left|\psi(t)\right\rangle }{\sqrt{\left\langle Q_{S}\right\rangle }} &  & p=k_{S}\left\langle Q_{S}\right\rangle dt\nonumber \\
\left|\psi(t+dt)\right\rangle =\left|T\right\rangle =\frac{J_{3}\left|\psi(t)\right\rangle }{\sqrt{\left\langle Q_{T}\right\rangle }} &  & p=k_{S}\left\langle Q_{T}\right\rangle dt\nonumber \\
\left|\psi(t+dt)\right\rangle =\left|T_{0}\right\rangle =\frac{J_{2}\left|\psi(t)\right\rangle }{\sqrt{\left\langle Q_{T}\right\rangle }} &  & p=k_{T}\left\langle Q_{T}\right\rangle dt\nonumber \\
\left|\psi(t+dt)\right\rangle =\left|S\right\rangle =\frac{J_{4}\left|\psi(t)\right\rangle }{\sqrt{\left\langle Q_{S}\right\rangle }} &  & p=k_{T}\left\langle Q_{S}\right\rangle dt
\end{eqnarray}

The form of the non-Hermitian effective Hamiltonian describing both
quantum measurement and chemical reaction (42) follows from (23) 
\begin{eqnarray}
H_{eff} & = & H-\frac{i}{2}\sum L_{i}^{+}L_{i}\nonumber \\
 & = & H-\frac{i}{2}\left(k_{S}Q_{S}+k_{T}Q_{S}+k_{S}Q_{T}+k_{T}Q_{T}\right)\nonumber \\
 & = & H-i\frac{k_{S}+k_{T}}{2}
\end{eqnarray}

The interpretation of (41) is that with probability $k_{S}dt$, the
wavefunction attempts to react via the singlet channel and with probability $k_{T}dt$, the wavefunction attempts to react
via the triplet channel. The singlet channel measures the wavefunction and with probability $k_{S}\left\langle Q_{S}\right\rangle dt$,
the wavefuction will react and form singlet product and with probability
$k_{S}\left\langle Q_{T}\right\rangle dt$ the wavefunction will be
{}``measured'' and forms a pure triplet, hence fails to react and
escapes the singlet recombination channel as a pure triplet. Analogously,
with probability $k_{T}\left\langle Q_{T}\right\rangle dt$ the wavefuction
will react and form triplet product and with probability $k_{T}\left\langle Q_{S}\right\rangle dt$
the wavefunction will be {}``measured'' and form a pure singlet.
Expanding (41) and keeping only first order terms in $dt$, one obtains

\begin{multline*}
\begin{array}{cc}
\rho(t+dt)= & \rho(t)-i[H,\rho]-\frac{k_{S}}{2}\{J_{1}^{+}J_{1},\rho(t)\}dt+\frac{k_{T}}{2}\{J_{2}^{+}J_{2},\rho(t)\}dt+k_{S}J_{1}\rho(t)J_{1}^{+}dt+k_{T}J_{2}\rho(t)J_{2}^{+}dt\\
 & -\frac{k_{S}}{2}(J_{3}\rho(t)+\rho(t)J_{3}-2J_{3}\rho(t)J_{3}^{+})dt-\frac{k_{T}}{2}(J_{4}\rho(t)+\rho(t)J_{4}-2J_{4}\rho(t)J_{4}^{+})dt
\end{array}
\end{multline*}
 projecting into the S-T basis and making use of the relation $J_{3}=1-J_{4}$,
equation (9) is recovered.

\subsection{Derivation of the Traditional Master Equation }

The phenomenological master equation can be derived in a way similar
to the Jones-Hore master equation, but using different jump operators. 

\begin{eqnarray}
\left|\psi(t+dt)\right\rangle =\frac{e^{-iH_{eff}dt}}{\sqrt{\left\langle \Psi\right|e^{iH_{eff}^{+}dt}e^{-iH_{eff}dt}\left|\Psi\right\rangle }}\left|\psi(t)\right\rangle  &  & p=1-k_{S}\left\langle Q_{S}\right\rangle dt-k_{T}\left\langle Q_{T}\right\rangle dt\nonumber \\
\left|\psi(t+dt)\right\rangle =\left|S_{0}\right\rangle =\frac{J_{1}\left|\psi(t)\right\rangle }{\sqrt{\left\langle Q_{S}\right\rangle }} &  & p=k_{S}\left\langle Q_{S}\right\rangle dt\nonumber \\
\left|\psi(t+dt)\right\rangle =\left|T_{0}\right\rangle =\frac{J_{2}\left|\psi(t)\right\rangle }{\sqrt{\left\langle Q_{T}\right\rangle }} &  & p=k_{T}\left\langle Q_{T}\right\rangle dt\label{eq:traditional}
\end{eqnarray}

\begin{equation}
H_{eff}=H-i\frac{k_{S}}{2}Q_{S}-i\frac{k_{T}}{2}Q_{T}
\end{equation}

The interpretation of (\ref{eq:traditional}) is that with probability
$k_{S}dt$, the wavefunction attempts to react via the singlet channel
and with probability $k_{T}dt$, the wavefunction attempts to react
via the triplet channel. With probabilities $k_{S}\left\langle Q_{S}\right\rangle dt$
and $ $$k_{T}\left\langle Q_{T}\right\rangle dt$ a reaction occurred.
However, \emph{nothing }can be said about reactants that attempt to
react but fail to do so. Therefore, a failed reaction doesn't return
a pure singlet or triplet \cite{Jones:2011wq}. 

In a qualitative sense, the difference between the two master equations
can be described using transition state theory. With a perfectly penetrative
barrier, reaching the barrier but not reacting necessarily implies
that one is in the wrong spin state, thus the fact that the molecule
is in the singlet channel yet does not react is physically significant.
With a partially reflective barrier, no reaction doesn't imply the
reactant is in the wrong spin state - reactant that is in the correct
spin state and reaches the barrier can still get reflected. 

\begin{figure}[H]
\subfloat[Fluorescence detection experiment ]{\includegraphics[scale=0.3]{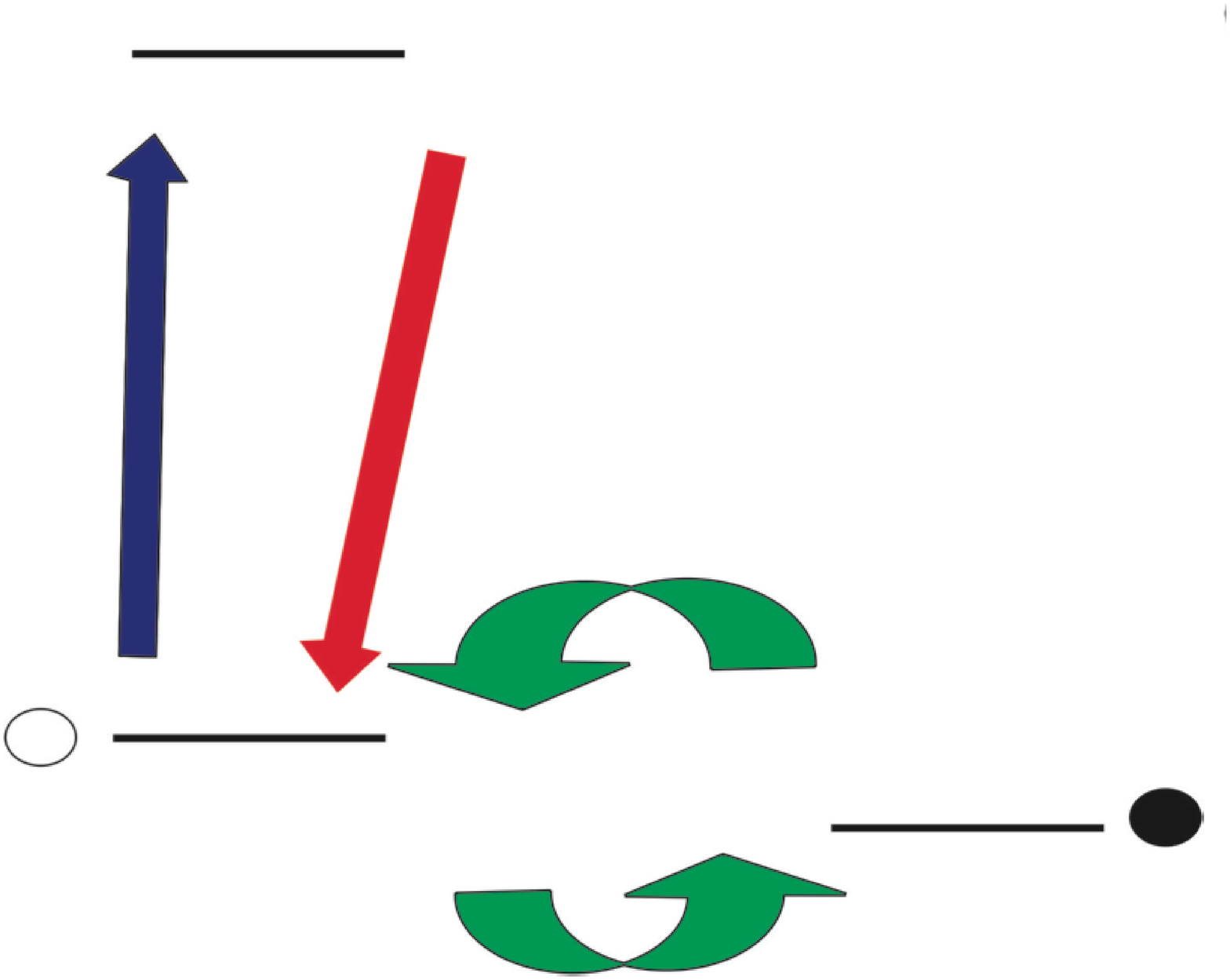}

} \subfloat[Spontaneous emission experiment ]{\includegraphics[scale=0.35]{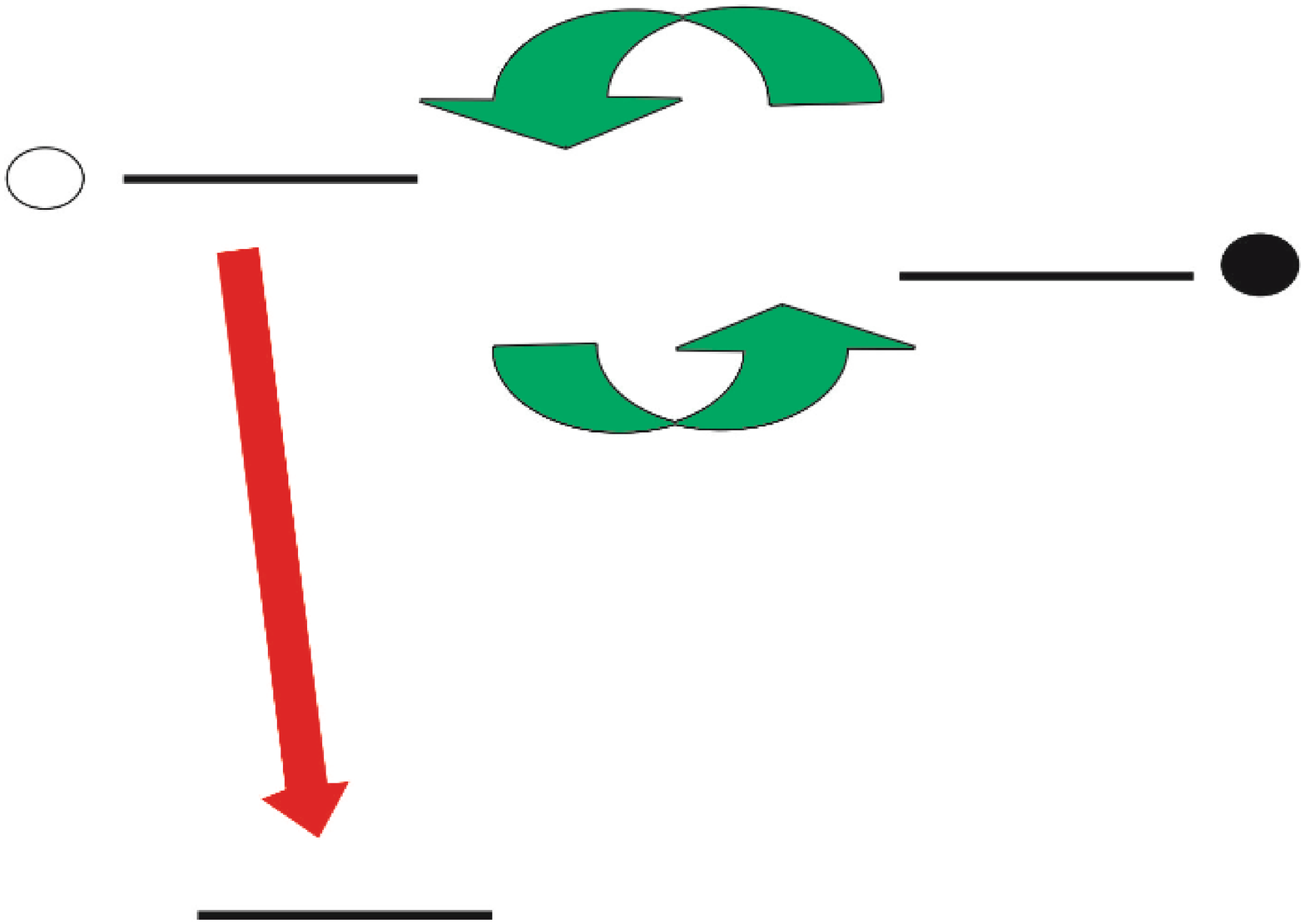}

}\caption{Analogy between a quantum optics experiment and a spin-selective reaction.
The blue arrow indicates excitation to an excited state, the red arrow
indicates radiative decay and the green arrow indicates coherent interconversion
between the energy states. The white and black spots indicates dark
and bright states respectively. \label{fig:Analogy-between-quantum}}
\end{figure}

Making the analogy between spin-selective chemical reaction and quantum
optics, the difference between the quantum measurement master equation
and the traditional master equation corresponds to measurement of
the state of a quantum optical system via a fluorescence detection
experiment or by observing a spontaneous decay process (Fig.\ref{fig:Analogy-between-quantum}).
In fluorescence detection \cite{Volz:2011zl}, light is shone on the
atom to excite it selectively from one of its two ground states, the
bright state, into a third excited state, whereupon it spontaneously
emits a photon and returns to the original state. The other ground
state, the dark state, is not excited by the incident light. Hence
if one observes no photon emitted after exciting the system, one is
sure that the system is in the dark state and thus null measurement
effectively collapses the wavefunction \cite{PORRATI:1987yg,QZE}. 

In the spontaneous emission scenario, two excited states are coherently
interconverting and only one state can decay to a ground state. If
one observes a photon, a measurement is performed and the wavefunction
collapsed onto the bright state before emitting a photon. However,
not observing a photon is not physically significant and does not
collapse the wavefunction. Of course, not observing a photon for infinitely
long time implies that the system is in the ground state, as explained
in the previous section. 

\begin{acknowledgments}
The author would like to thank Professor Peter Hore and Professor
Jonathan Jones for their expert advice. This work has been supported
by a Nuffield Undergraduate Research Bursary. 
\end{acknowledgments}
\bibliographystyle{ChemCommun}
\addcontentsline{toc}{section}{\refname}\bibliography{Kominis}

\appendix

\section{Microscopic derivation of the Jones-Hore Master Equation}

Following Carmichael \cite{caramichael1}, the total Hamiltonian is
the sum of system, environment and interaction Hamiltonians: 

\begin{equation}
H=H_{sys}+H_{env}+H_{int}
\end{equation}

Moving to the interaction picture, the Liouville-von Neuman equation
is 

\begin{equation}
\frac{d\widetilde{\rho}}{dt}=-i[\tilde{H}_{int},\widetilde{\rho}]
\end{equation}

Formal integration gives 
\begin{equation}
\widetilde{\rho}(t)=\widetilde{\rho}(0)-i\intop_{0}^{t}dt'[\tilde{H}_{int}(t'),\widetilde{\rho}(t')]
\end{equation}

Substituting $\widetilde{\rho}(t)$ back into the right hand side
of (A2) gives 
\begin{equation}
\frac{d\widetilde{\rho}}{dt}=-i[\tilde{H}_{int}(t),\widetilde{\rho}(0)]-\intop_{0}^{t}dt'[\tilde{H}_{int}(t),[\tilde{H}_{int}(t'),\widetilde{\rho}(t')]]
\end{equation}

Tracing out the reservoir degrees of freedom, and assuming that there
are no initial system-reservoir correlations, i.e $tr_{R}([\tilde{H}_{int}(t),\widetilde{\rho}(0)])$
\begin{equation}
\frac{d\widetilde{\rho}_{S}(t)}{dt}=-\intop_{0}^{t}dt'tr_{R}\left\{ [\tilde{H}_{int}(t),[\tilde{H}_{int}(t'),\widetilde{\rho}(t')]]\right\} 
\end{equation}
where $\widetilde{\rho}_{S}=tr_{R}(\rho)$ is the density matrix of
the system. In the limit of weak coupling and an infinitely large
reservoir, the motion of the system and reservoir remains factored
throughout the evolution, hence 
\begin{equation}
\widetilde{\rho}(t)=R_{0}\widetilde{\rho}_{S}(t)
\end{equation}
where $R_{0}$ is the initial reservoir state. Substituting in (A5)
gives 
\begin{equation}
\frac{d\widetilde{\rho}_{S}(t)}{dt}=-\intop_{0}^{t}dt'tr_{R}\left\{ [\tilde{H}_{int}(t),[\tilde{H}_{int}(t'),R_{0}\widetilde{\rho}_{S}(t')]]\right\} 
\end{equation}

Writing 
\begin{equation}
\tilde{H}_{int}=\sum_{i}\tilde{\Gamma}_{i}\widetilde{s}_{i}
\end{equation}
where $\widetilde{\Gamma_{i}}$ are operators belonging to the reservoir
subspace and $\widetilde{s}_{i}$ are operators belonging to the system
space, gives 
\begin{equation}
\frac{d\widetilde{\rho}_{S}(t)}{dt}=-\sum_{i,j}\intop_{0}^{t}dt'tr_{R}\left\{ [\tilde{\Gamma}_{i}(t)\widetilde{s}_{i}(t),[\tilde{\Gamma}_{j}(t')\widetilde{s}_{j}(t'),R_{0}\widetilde{\rho}_{S}(t')]]\right\} 
\end{equation}

Expanding out the sum and recognizing reservoir correlation functions,
we obtain 
\begin{equation}
\frac{d\widetilde{\rho}_{S}(t)}{dt}=-\sum_{i,j}\intop_{0}^{t}dt'[\widetilde{s_{i}}(t)\widetilde{s_{j}}(t')\widetilde{\rho}_{S}(t')-\widetilde{s_{j}}(t')\widetilde{\rho}_{S}(t')\widetilde{s_{i}}(t)]\left\langle \tilde{\Gamma}_{i}(t)\tilde{\Gamma}_{j}(t')\right\rangle _{R}+h.c
\end{equation}
where
\begin{equation}
\left\langle \tilde{\Gamma}_{i}(t)\tilde{\Gamma}_{j}(t')\right\rangle _{R}=tr_{R}[R_{0}\tilde{\Gamma}_{i}(t)\tilde{\Gamma}_{j}(t')]
\end{equation}
 The system Hamiltonian, in the $\{S,T,S_{0},T_{0}\}$ basis, is 
\begin{equation}
H_{sys}=\left(\begin{array}{cccc}
\omega_{0} & 0 & 0 & 0\\
0 & \omega_{1} & 0 & 0\\
0 & 0 & 0 & 0\\
0 & 0 & 0 & 0
\end{array}\right)=\omega_{0}J_{1}^{+}J_{1}+\omega_{1}J_{2}^{+}J_{2}
\end{equation}

The environment is assumed to be a series of harmonic oscillators
that absorbs the energy dissipated from the system and two measurement
devices, which are also modeled as harmonic oscillators 
\begin{equation}
H_{env}=\sum_{k}\omega_{k}a_{k,1}^{+}a_{k,1}+\sum_{k}\omega_{k}a_{k,2}^{+}a_{k,2}+\omega_{b}\sum_{k}b_{k}^{+}b_{k}+\omega_{c}\sum_{k}c_{k}^{+}c_{k}
\end{equation}

The interaction Hamiltonian comprises two parts, a dissipative part
($H_{diss})$ which is modeled as the Jaynes-Cumming Hamiltonian \cite{caramichael1}
and a measurement Hamiltonian ($H_{mea}$) \cite{puredepha}. 
\begin{equation}
H_{diss}=\sum_{k}(g_{k}a_{k,1}^{+}J_{1}+g_{k}^{*}a_{k,1}J_{1}^{+})+\sum_{k'}(q_{k'}a_{k',2}^{+}J_{2}+q_{k'}^{*}a_{k',2}J_{2}^{+})
\end{equation}
\begin{equation}
H_{mea}=J_{1}^{+}J_{1}\sum_{k}(\alpha_{k}b_{k}^{+}+\alpha_{k}^{*}b_{k})+J_{2}^{+}J_{2}\sum_{k'}(\beta_{k'}c_{k'}^{+}+\beta_{k'}^{*}c_{k'})
\end{equation}

The measurement device $b$ measures the singlet occupancy and the
device $c$ measures the triplet occupancy. Note that $[H_{sys},H_{mea}]=0$
and $[H_{sys},H_{diss}]\neq0$, implying that energy is conserved
by the measurement process but not conserved by the dissipative dynamics.
Looking at the form of the measurement Hamiltonian, at long times,
the measurement devices $b$ and $c$ will collapse to pointer states
$J_{1}^{+}J_{1}$ and $J_{2}^{+}J_{2}$$ $ respectively \cite{quanstatmech}.
Let 
\begin{equation}
\widetilde{\Gamma}_{A,1}=\sum_{k}g_{k}^{*}a_{k,1}e^{-i\omega_{k}t}
\end{equation}

\begin{equation}
\widetilde{\Gamma}_{A,2}=\sum_{k}q_{k}^{*}a_{k,2}e^{-i\omega_{k}t}
\end{equation}

\begin{equation}
\widetilde{\Gamma}_{B}=\sum_{k}\alpha_{k}^{*}b_{k}e^{-i\omega_{k}t}
\end{equation}

\begin{equation}
\widetilde{\Gamma}_{C}=\sum_{k}\beta_{k}^{*}c_{k}e^{-i\omega_{k}t}
\end{equation}

Writing out explicitly the operators in the form of (A8) 
\[
\widetilde{\Gamma}_{1}=\widetilde{\Gamma}_{A,1}^{+}\qquad\widetilde{s}_{1}=J_{1}e^{-i\omega_{0}t}
\]
\[
\widetilde{\Gamma}_{2}=\widetilde{\Gamma}_{A,1}\qquad\widetilde{s}_{2}=J_{1}^{+}e^{i\omega_{0}t}
\]

\[
\widetilde{\Gamma}_{3}=\widetilde{\Gamma}_{A,2}^{+}\qquad\widetilde{s}_{3}=J_{2}e^{-i\omega_{1}t}
\]
\[
\widetilde{\Gamma}_{4}=\widetilde{\Gamma}_{A,2}\qquad\widetilde{s}_{4}=J_{2}^{+}e^{i\omega_{1}t}
\]
\[
\widetilde{\Gamma}_{5}=\widetilde{\Gamma}_{B}+\widetilde{\Gamma}_{B}^{+}\qquad\widetilde{s}_{5}=J_{1}^{+}J_{1}
\]
\[
\widetilde{\Gamma}_{6}=\widetilde{\Gamma}_{C}+\widetilde{\Gamma}_{C}^{+}\qquad\widetilde{s}_{6}=J_{2}^{+}J_{2}
\]

At zero temperature, reservoirs $A_{1}$, $A_{2},$ B and C have the
following correlation functions 
\begin{equation}
\left\langle \tilde{\Gamma}(t)\tilde{\Gamma}(t')\right\rangle _{R}=0
\end{equation}
\begin{equation}
\left\langle \tilde{\Gamma}^{+}(t)\tilde{\Gamma}^{+}(t')\right\rangle _{R}=0
\end{equation}
\begin{equation}
\left\langle \tilde{\Gamma}^{+}(t)\tilde{\Gamma}(t')\right\rangle _{R}=0
\end{equation}
\begin{equation}
\left\langle \tilde{\Gamma}(t)\tilde{\Gamma}^{+}(t')\right\rangle _{R}=\sum_{j}\left|k_{j}\right|^{2}e^{-i\omega_{j}(t-t')}=\intop_{0}^{\infty}d\omega\left|k(\omega)\right|^{2}g(\omega)e^{-i\omega(t-t')}
\end{equation}
where $k_{j}$ is the generic system-reservoir coupling constant,
and $g(\omega)$ is the density of states in the continuum limit.
Substituting and noting that the $S$ and $T$ states are orthogonal
and reservoirs $A_{1}$, $A_{2},$ B and C are uncorrelated and thus
statistically independent 

\begin{eqnarray}
\frac{d\widetilde{\rho}_{S}(t)}{dt}= & \intop_{0}^{t}dt'[J_{1}^{+}J_{1}\widetilde{\rho}_{S}(t')-J_{1}\widetilde{\rho}_{S}(t')J_{1}^{+}]e^{-i\omega_{0}(t-t')}\left\langle \tilde{\Gamma}_{A,1}(t)\tilde{\Gamma}_{A,1}^{+}(t')\right\rangle _{R}\nonumber \\
 & +[J_{2}^{+}J_{2}\widetilde{\rho}_{S}(t')-J_{2}\widetilde{\rho}_{S}(t')J_{2}^{+}]e^{-i\omega_{1}(t-t')}\left\langle \tilde{\Gamma}_{A,2}(t)\tilde{\Gamma}_{A,2}^{+}(t')\right\rangle _{R}\nonumber \\
 & +[J_{1}^{+}J_{1}\widetilde{\rho}_{S}(t')-J_{1}^{+}J_{1}\widetilde{\rho}_{S}(t')J_{1}^{+}J_{1}]\left\langle \tilde{\Gamma}_{B}(t)\tilde{\Gamma}_{B}^{+}(t')\right\rangle _{R}\nonumber \\
 & +[J_{2}^{+}J_{2}\widetilde{\rho}_{S}(t')-J_{2}^{+}J_{2}\widetilde{\rho}_{S}(t')J_{2}^{+}J_{2}]\left\langle \tilde{\Gamma}_{C}(t)\tilde{\Gamma}_{C}^{+}(t')\right\rangle _{R}+h.c
\end{eqnarray}

Substituting $\tau=t-t'$

\begin{eqnarray}
\frac{d\widetilde{\rho}_{S}(t)}{dt}= & \intop_{0}^{t}d\tau[J_{1}^{+}J_{1}\widetilde{\rho}_{S}(t-\tau)-J_{1}\widetilde{\rho}_{S}(t-\tau)J_{1}^{+}]e^{-i\omega_{0}\tau}\left\langle \tilde{\Gamma}_{A,1}(t)\tilde{\Gamma}_{A,1}^{+}(t-\tau)\right\rangle _{R}\nonumber \\
 & +[J_{2}^{+}J_{2}\widetilde{\rho}_{S}(t-\tau)-J_{2}\widetilde{\rho}_{S}(t-\tau)J_{2}^{+}]e^{-i\omega_{1}\tau}\left\langle \tilde{\Gamma}_{A,2}(t)\tilde{\Gamma}_{A,2}^{+}(t-\tau)\right\rangle _{R}\nonumber \\
 & +[J_{1}^{+}J_{1}\widetilde{\rho}_{S}(t-\tau)-J_{1}^{+}J_{1}\widetilde{\rho}_{S}(t-\tau)J_{1}^{+}J_{1}]\left\langle \tilde{\Gamma}_{B}(t)\tilde{\Gamma}_{B}^{+}(t-\tau)\right\rangle _{R}\nonumber \\
 & +[J_{2}^{+}J_{2}\widetilde{\rho}_{S}(t-\tau)-J_{2}^{+}J_{2}\widetilde{\rho}_{S}(t-\tau)J_{2}^{+}J_{2}]\left\langle \tilde{\Gamma}_{C}(t)\tilde{\Gamma}_{C}^{+}(t-\tau)\right\rangle _{R}+h.c
\end{eqnarray}

Performing the Markov approximation and replacing $t-\tau$ terms
in the density operator by $t$
\begin{eqnarray}
\frac{d\widetilde{\rho}_{S}(t)}{dt}= & [J_{1}^{+}J_{1}\widetilde{\rho}_{S}(t)-J_{1}\widetilde{\rho}_{S}(t)J_{1}^{+}]\intop_{0}^{t}d\tau e^{-i\omega_{0}\tau}\left\langle \tilde{\Gamma}_{A,1}(t)\tilde{\Gamma}_{A,1}^{+}(t-\tau)\right\rangle _{R}\nonumber \\
 & +[J_{2}^{+}J_{2}\widetilde{\rho}_{S}(t)-J_{2}\widetilde{\rho}_{S}(t)J_{2}^{+}]\intop_{0}^{t}d\tau e^{-i\omega_{1}\tau}\left\langle \tilde{\Gamma}_{A,2}(t)\tilde{\Gamma}_{A,2}^{+}(t-\tau)\right\rangle _{R}\nonumber \\
 & +[J_{1}^{+}J_{1}\widetilde{\rho}_{S}(t)-J_{1}^{+}J_{1}\widetilde{\rho}_{S}(t)J_{1}^{+}J_{1}]\intop_{0}^{t}d\tau\left\langle \tilde{\Gamma}_{B}(t)\tilde{\Gamma}_{B}^{+}(t-\tau)\right\rangle _{R}\nonumber \\
 & +[J_{2}^{+}J_{2}\widetilde{\rho}_{S}(t)-J_{2}^{+}J_{2}\widetilde{\rho}_{S}(t)J_{2}^{+}J_{2}]\intop_{0}^{t}d\tau\left\langle \tilde{\Gamma}_{C}(t)\tilde{\Gamma}_{C}^{+}(t-\tau)\right\rangle _{R}+h.c.
\end{eqnarray}

Ignoring imaginary frequency shifts, one can define

\begin{equation}
\intop_{0}^{t}d\tau e^{-i\omega_{0}\tau}\left\langle \tilde{\Gamma}_{A,1}(t)\tilde{\Gamma}_{A,1}^{+}(t-\tau)\right\rangle _{R}=\intop_{0}^{t}d\tau\left\langle \tilde{\Gamma}_{B}(t)\tilde{\Gamma}_{B}^{+}(t-\tau)\right\rangle _{R}=\frac{k_{S}}{2}
\end{equation}

\begin{equation}
\intop_{0}^{t}d\tau e^{-i\omega_{1}\tau}\left\langle \tilde{\Gamma}_{A,2}(t)\tilde{\Gamma}_{A,2}^{+}(t-\tau)\right\rangle _{R}=\intop_{0}^{t}d\tau\left\langle \tilde{\Gamma}_{C}(t)\tilde{\Gamma}_{C}^{+}(t-\tau)\right\rangle =\frac{k_{T}}{2}
\end{equation}

\begin{eqnarray}
\frac{d\widetilde{\rho}_{S}(t)}{dt} & = & -\frac{k_{S}}{2}(J_{1}^{+}J_{1}\widetilde{\rho}_{S}(t)+\widetilde{\rho}_{S}(t)J_{1}^{+}J_{1}-2J_{1}\widetilde{\rho}_{S}(t)J_{1}^{+})-\frac{k_{T}}{2}\left(J_{2}^{+}J_{2}\widetilde{\rho}_{S}(t)-\widetilde{\rho}_{S}(t)J_{2}^{+}J_{2}-2J_{2}\widetilde{\rho}_{S}(t)J_{2}^{+}\right)\nonumber \\
 &  & -\frac{k_{s}+k_{T}}{2}\left(J_{1}^{+}J_{1}\widetilde{\rho}_{S}(t)+\widetilde{\rho}_{S}(t)J_{1}^{+}J_{1}-2J_{1}^{+}J_{1}\widetilde{\rho}_{S}(t)J_{1}^{+}J_{1}\right)
\end{eqnarray}

Converting back to the Schrödinger picture and projecting into \{S,T\}
subspace, (9) is recovered.

\section*{}
\end{document}